# Optical wavelength conversion via optomechanical coupling in a silica resonator

Chunhua Dong[1], Victor Fiore[1], Mark C. Kuzyk[1], Lin Tian[2], and Hailin Wang[1]

[1]Department of Physics, University of Oregon, Eugene, Oregon 97403, USA
[2]5200 North Lake Road, University of California, Merced, California 95343, USA

Abstract

We report the experimental demonstration of converting coherent optical fields between two different optical wavelengths by coupling two optical modes to a mechanical breathing mode in a silica resonator. The experiment is based on an itinerant approach, in which state-mapping from optical to mechanical and from mechanical to another optical state takes place simultaneously. In contrast to conventional nonlinear optical processes, optomechanical impedance matching as well as efficient optical input-output coupling, instead of phase-matching, plays a crucial role in optomechanics-based wavelength conversion.

Recent studies of interactions between an optical and a mechanical mode via radiation pressure force in an optomechanical resonator have led to the experimental realization of a number of remarkable phenomena[1,2], including the cooling of a mechanical mode to its motional ground state[3-7], optomechanically-induced transparency[8-11], the strong coupling between optical and mechanical modes[11-13], and optomechanical light storage[14]. These advances have opened up avenues for taking advantage of unique properties of optomechanical processes for quantum information processing and for optomechanics-based nonlinear optics.

In an optomechanical resonator, an optically-active mechanical mode can couple to any of the optical resonances via radiation pressure. This unique property can enable an important application: state transfer of photons with vastly different wavelengths[15-20]. Such state transfer plays an essential role in a quantum network. For example, an optical field can be converted to a wavelength that is suitable for long distance transport or to wavelengths that can couple to different quantum systems, such as trapped ions, quantum dots, diamond color-centers, or superconducting circuits. A number of approaches for optical wavelength conversion in an optomechanical resonator have been proposed[15-20]. In all these approaches, two optical modes couple to a common mechanical mode via radiation pressure forces. In a double-swap approach, successive optomechanical $\pi/2$-pulses drive two state-mapping processes[16]. An optical field in the first optical mode is mapped to a mechanical excitation[21]. The mechanical excitation is then mapped to an optical field in the second optical mode. In an itinerant approach, the two state-mapping processes take place simultaneously. Recently, adiabatic transfer of optical states via a dark mechanical mode has also been proposed[19,20]. This adiabatic approach can in principle be immune to mechanical dissipation, eliminating contributions of thermal phonons to the state-mapping process.

Here, we report an experimental demonstration of optomechanics-based wavelength conversion of coherent optical fields. We have realized the itinerant approach in a silica optomechanical resonator, demonstrating the feasibility of exploiting optomechanical processes for state-transfer of photons between vastly different wavelengths. We have also investigated the behavior of the mechanical mode involved in the wavelength conversion. These studies show that optomechanical impedance matching as well as efficient optical input-output coupling, instead of phase matching, plays a crucial role in optomechanics-based wavelength conversion. The optomechanical impedance matching ensures that the state transfer from photon to phonon and then to photon can proceed without excessive radiation-pressure damping of the mechanical excitation.

Figure 1a illustrates an optomechanical system, in which two optical cavity modes couple to a mechanical oscillator. The optomechanical coupling is driven by two strong laser fields, $E_1$ and $E_2$, near the red sideband of the respective optical modes (i.e., one mechanical frequency, $\omega_m$, below cavity resonances, $\omega_1$ and $\omega_2$), as shown in Fig. 1b. In the resolved-sideband limit, the linearized optomechanical Hamiltonian is given by[19, 20]

$$H = \hbar\omega_m \hat{b}^+\hat{b} - \hbar(\Delta_1 \hat{a}_1^+\hat{a}_1 + \Delta_2 \hat{a}_2^+\hat{a}_2) + \hbar G_1(\hat{a}_1^+ b + \hat{a}_1 b^+) + \hbar G_2(\hat{a}_2^+ b + \hat{a}_2 b^+) \qquad (1)$$

where $\hat{b}$ is the annihilation operator for the mechanical mode, $\hat{a}_1$ and $\hat{a}_2$ are the annihilation operators for the optical signal fields in the respective rotating frames of the external driving fields, $\Delta_1$ and $\Delta_2$ are the detuning between the external driving field and the relevant cavity resonance, and $G_1$ and $G_2$ are the effective optomechanical coupling rates between the respective optical signal field and the mechanical displacement.

For the optical wavelength conversion, the optomechanical coupling characterized by $G_1$ maps an input signal field, $E_{in}$, to a mechanical excitation. The coupling characterized by $G_2$ maps the mechanical excitation to an output signal field, $E_{out}$. We assume that $E_{in}$ is near $\omega_1$ and $E_{out}$ is near $\omega_2$. In the steady state, the photon conversion efficiency, which is defined as the ratio of the output-signal photon flux over the input-signal photon flux, is given by,

$$\chi = \eta_1 \eta_2 \frac{4 C_1 C_2}{(1 + C_1 + C_2)^2} \qquad (2)$$

where $C_i = 4G_i^2 / \gamma_m \kappa_i$ and $\eta_i = \kappa_i^{ext} / \kappa_i$ ($i = 1, 2$) are the optomechanical cooperativity and the output coupling ratio for the two optical modes, respectively, $\kappa_i$ is the total cavity decay rate, $\kappa_i^{ext}$ is the effective output coupling rate, and $\gamma_m$ is the mechanical damping rate. We also assumed that $\Delta_1 = \Delta_2 = -\omega_m$ and the input and output signal fields are resonant with the respective optical cavity mode. Near unity photon conversion can be achieved in the limit that $C_1 = C_2 \gg 1$ and that there is no optical loss other than the input-output coupling[19, 20].

We have used silica microspheres with a diameter near 30 μm as a model optomechanical resonator[22]. Silica resonators including spheres and toroids can feature whispering gallery modes (WGMs) with ultrahigh optical finesse in a broad wavelength range[23].

Figure 2a shows a schematic of the experimental setup (see the Supplement for a detailed setup), where we chose an input signal field with $\lambda_1 \sim 800$ nm, which is near the wavelengths for optical transitions in semiconductors such as GaAs, and an output signal field with $\lambda_2 \sim 637$ nm, which is near the wavelength for the zero-phonon optical transition of nitrogen vacancy centers in diamond. The frequencies of the two driving fields are fixed with $\Delta_1 = \Delta_2 = -\omega_m$. Examples of WGM transmission resonances obtained near 800 nm and 637 nm are shown in Fig. 2b. The two WGMs are coupled to the (1, 2) mechanical breathing mode of the microsphere. Figures 2b and 2c show the displacement power spectrum and the spatial displacement pattern of the (1, 2) mode, respectively. Optical and mechanical parameters of the silica resonators used include $(\kappa_1, \kappa_2)/2\pi \approx 30$ MHz, $\omega_m/2\pi \approx 101$ MHz, and $\gamma_m/2\pi \approx 20$ kHz.

To avoid heating induced by the strong driving fields, we kept the duty cycle of the experiment to below 5% by using relatively long optical pulses for the wavelength conversion experiment. The input signal pulse and the two driving pulses were generated by gating the continuous-wave laser beams with acousto-optical modulators. The three optical pulses were then coupled into a single mode fiber. All three pulses have the same timing and equal duration, with a pulse duration of 6 µs unless otherwise specified. The output signal was measured with heterodyne-detection, for which $E_2$ serves as the local oscillator. The power density spectrum of the heterodyne signal was obtained in a given time window, with a spectrum analyzer operating in a gated-detection mode[14].

Figure 3 shows the photon conversion efficiency obtained as a function of the input power, $P_1$, for $E_1$ and at various fixed power, $P_2$, for $E_2$ (see the Supplement for a detailed discussion on the calibration of the output signal power). The data were obtained with the detection gate set between 4 and 5 µs of the signal pulse. Optical powers, including $P_1$, $P_2$, and input signal power, $P_{in}$, were measured in front of the fiber facet. Figure 3 features a distinct saturation behavior. The conversion efficiency increases and then saturates with increasing $P_1$, with the saturation power depending on $P_2$. More detail analysis will be presented later.

For the transient wavelength conversion, both the amplitude and the spectral response of the output signal depend on the duration of the conversion process. Figure 4a plots the output signal power as a function of time (or gate delay), obtained with a detection gate length of 0.5 µs and with $P_1 = 25$ mW and $P_2 = 6$ mW. For comparison, Fig. 4a also shows the data obtained under the same condition except that the pulse duration is 3 µs. For these experiments, the output signal power reaches steady state in about 3 µs. Note that in the limit of small $G_1$ and $G_2$, the output signal power reaches the steady state in a timescale of the mechanical damping time.

Figure 4b shows the dependence of the output signal power on the detuning, δ, between the input signal and $E_1$, obtained with pulse durations of 6 µs as well as 3 µs. The output signal was measured with a gate length of 1 µs and with the detection gate centered in the second half of the signal pulse. As shown in Fig. 4b, a sharp resonance occurs when the detuning is near $\omega_m$. The linewidth of the resonance decreases with increasing pulse duration. We also examined the dependence of the linewidth on the power of the two driving fields. No discernable power dependence was observed for the relatively small $P_1$ and $P_2$ used in the experiment.

We used a weak 3 µs pulse to probe the mechanical mode involved in the wavelength conversion process. Gated detection was also used for the measurement of the displacement power density spectrum of the mechanical mode. The gate length was set to 1 µs and the gate was at the center of the probe pulse. The probe pulse is at the same frequency as $E_1$ and arrives 1 µs after $E_1$. Since the mechanical damping time is much longer than the optical cavity lifetime, the 1 µs delay is sufficiently long for the optical fields involved in the conversion process to escape from the resonator, but is short enough to get a snap shot on the behavior of the mechanical mode before its eventual thermalization. Figure 5a shows the intensity of the mechanical oscillation as a function of $P_1$, with $P_2$=0. Figure 5b shows the intensity of the mechanical oscillation measured by the probe as a function of $P_2$, with $P_1$=1 mW. The intensity of the mechanical oscillation was derived from the spectrally integrated area of the displacement power spectrum obtained with the gated detection.

The experimental results shown in Fig. 5 provide important insights on how the mechanical motion contributes to the optical wavelength conversion and also how the conversion process affects the mechanical motion. As shown in Fig. 5a, the mechanical intensity increases with $P_1$ at relatively small $P_1$. In this case, $E_1$ and $E_{in}$ interact with the mechanical mode, generating a mechanical oscillation and effectively mapping the input signal field at $\omega_l$ to a coherent mechanical excitation[14]. At relatively large $P_1$, the mechanical intensity, however, decreases with increasing $P_1$, indicating that the strong optomechanical interaction not only performs the optical-to-motional state transfer, but also damps the mechanical oscillation.

Note that $E_1$ always damps or cools down the thermal mechanical motion[24, 25].

Figure 5b shows that the mechanical intensity decreases monotonically with $P_2$. In this case, $E_2$ couples to the coherent mechanical excitation and maps the mechanical excitation to an output signal field at $\omega_2$, which effectively damps the coherent mechanical excitation. It should be noted that the interaction between $E_2$ and the thermal mechanical motion can also generate a thermal background in the output signal (see the Supplement). With a relatively strong coherent input signal, the thermal background can be negligibly small. The thermal background, however, becomes important for conversion of single photons or nonclassical optical states. In this regard, the effective cooling of the thermal mechanical motion by $E_1$ can be highly beneficial for extending the wavelength conversion process to the quantum regime.

For a detailed theoretical analysis of the experimental results, we have used coupled oscillator equations to describe the coupling between the mechanical displacement and the two optical modes (see the Supplement), with nearly all parameters, including optical and mechanical linewidths, determined from the experiment. Since it is difficult to accurately estimate the optomechanical coupling rate from the input powers of the two driving fields, we used as adjustable parameters two proportional constants that relate the input optical power, $P_1$ and $P_2$, to the optomechanical cooperativity, $C_1$ and $C_2$. We determined the proportional constants from the dependence of the mechanical intensity on $P_1$ and $P_2$ shown in Fig. 5. The same proportional constants were then used for the calculations shown in Figs. 3 and 4.

Figure 5 compares the experimentally obtained mechanical intensity as a function of $P_1$ and $P_2$ with the corresponding theoretical calculation. Good agreement between the theory and experiment is obtained with $P_1 = 5$ mW and $P_2 = 15$ mW corresponding to $C_1 = 1$ and $C_2 = 1$, respectively, which sets the two proportional constants discussed above. Much higher $P_2$ was needed for achieving $C_2 = 1$ than $P_1$ for achieving $C_1 = 1$, in part because for our setup, the efficiency of coupling a laser beam with $\lambda \sim 637$ nm into the single mode fiber is considerably smaller than that with $\lambda \sim 800$ nm. Figure 4b shows the good agreement between the theory and experiment in terms of the spectral response of the wavelength conversion process. All parameters used, other than the proportional constants determined from Fig. 5, were determined directly from the experiment.

The experimentally obtained conversion efficiency is also in good agreement with the theoretical calculation, as shown in Fig. 3. The parameters used for the numerical calculation include the optical and mechanical linewidths, the proportional constants derived from Fig. 5, and estimated output coupling ratio, $\eta_1\eta_2=(0.45)^2$ for data obtained with $P_2=2$ mW and $P_2=11$ mW and $\eta_1\eta_2=(0.5)^2$ for data obtained with $P_2=21$ mW. The relatively small conversion efficiency observed is in part due to the modest optomechanical cooperativity used in the experiment. Optical loss other than input-output coupling is also a primary limiting factor. To achieve greater conversion efficiency, we increased the coupling of the incident optical fields to the WGMs by reducing slightly the gap between the fiber taper and the microsphere for the experiment with $P_2=21$ mW. Significantly greater conversion efficiency can be achieved by further increasing the output coupling ratio and operating the experiment in an over-coupled regime.

The distinct saturation behavior shown in Fig. 3 reflects the requirement of optomechanical impedance matching. As shown in Fig. 5, the two driving fields, $E_1$ and $E_2$, not only perform the state mapping between optical and motional states, but also damp the mechanical excitation. In the limit of $C_1 \gg C_2 > 1$, $E_1$ leads to excessive damping of the mechanical excitation and prevents its efficient conversion into the output signal by $E_2$. In the limit of $C_2 \gg C_1 > 1$, $E_2$ leads to excessive mechanical damping, hindering the efficient generation of a coherent mechanical excitation by $E_1$. As shown in Eq. 2, efficient wavelength conversion requires not only large, but also equal optomechanical cooperativity for the two optical modes. The impedance matching condition, $C_1=C_2$, ensures that the state transfer from photon to phonon and then to photon can proceed without excessive damping of the mechanical excitation by either $E_1$ or $E_2$. In comparison, for a convention nonlinear optical process[26], phase matching of the optical waves involved is an essential requirement, which limits the spectral range of the nonlinear optical process.

In order to extend the wavelength conversion process to the quantum regime, it becomes necessary to suppress contributions from thermal phonons. This can be accomplished by cooling the mechanical mode to its motional ground state or alternatively by using adiabatic state transfer through a dark mechanical mode[19, 20]. To further improve the conversion efficiency, one can also exploit nanostructures such as optomechanical crystals that feature greater optomechanical cooperativity[10]. While the conversion efficiency is ultimately limited by optical losses other than input-output coupling, the photon losses can in principle be compensated with error correction schemes[27].

This work is supported by DARPA-MTO through a grant from AFOSR.

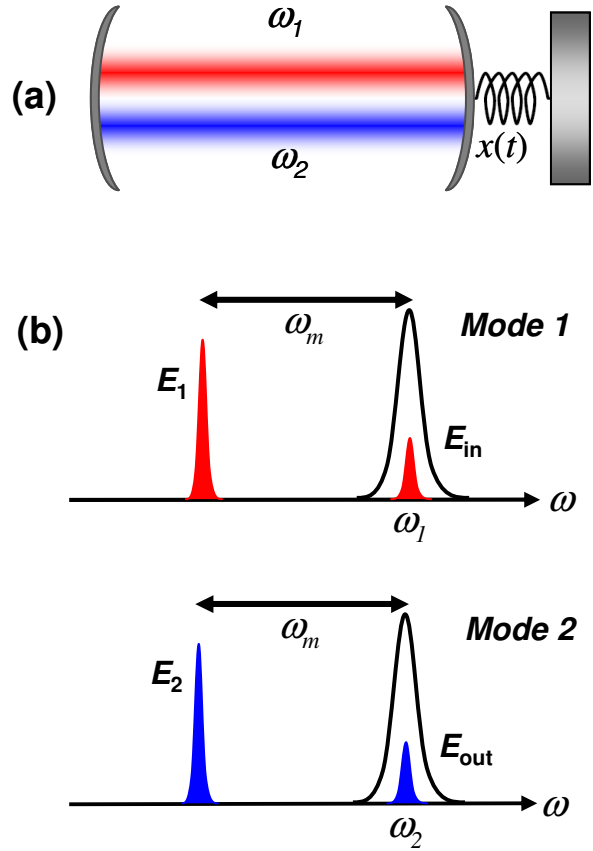

FIG 1. (color on line). (a) Schematic of two optical modes coupling to a mechanical oscillator. (b) Optical fields, $E_1$ and $E_2$, at the red sideband of the respective optical resonance drive the coupling between the mechanical excitation and the optical signal fields, resulting in a state transfer between input signal field, $E_{in}$, and output signal field, $E_{out}$.

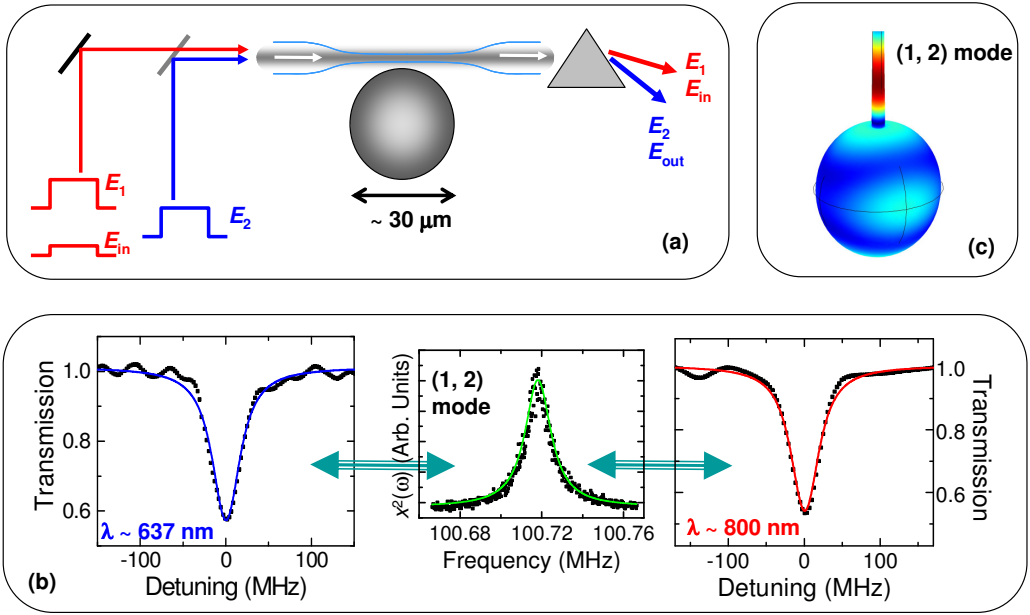

FIG 2. (color on line). (a) A simplified schematic of the experimental setup. A tapered single mode fiber is used for input and output coupling of WGMs in the silica microsphere. (b) Examples of transmission resonances of two WGMs, along with the displacement power spectrum of the (1, 2) mechanical mode. Solid lines are fit to Lorentzian lineshapes. (c) Calculated spatial displacement pattern of the (1, 2) mode.

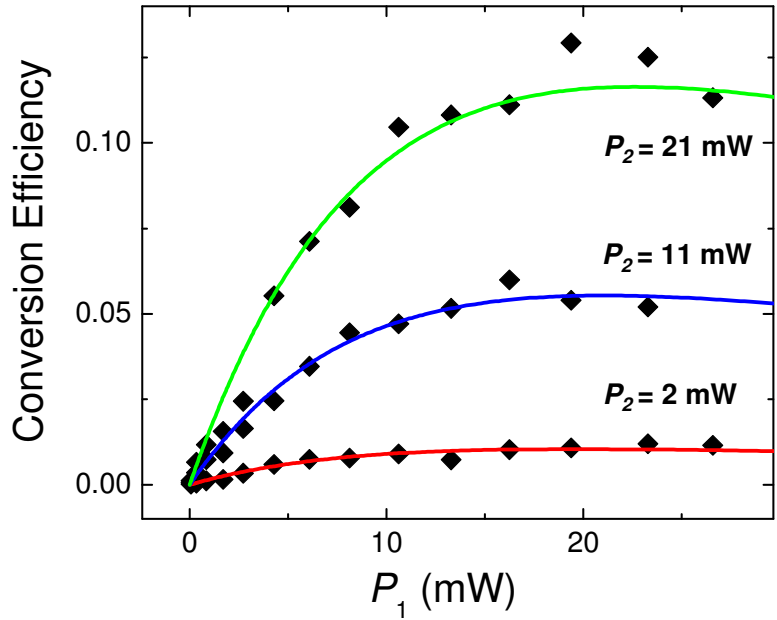

FIG 3. (color on line). Photon conversion efficiency. The signal field is converted from $\lambda \sim 800$ nm to $\lambda \sim 637$ nm, with input signal power, $P_{in}$=0.2 mW. $P_1$ and $P_2$ are the optical power of the driving fields with wavelength near 800 nm and 637 nm, respectively. The solid lines are the theoretically calculated conversion efficiency, as discussed in the text.

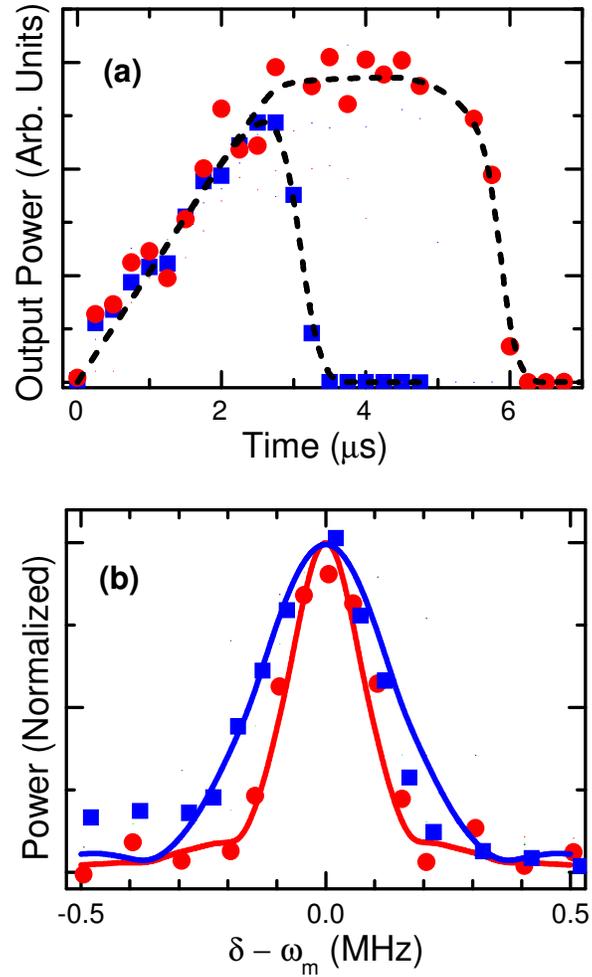

FIG 4. (color on line). (a) The output signal power as a function of time, obtained with $P_1$=25 mW and $P_2$=6 mW. The dashed lines are a guide to the eye. (b) The output signal power as a function of the detuning, $\delta$, between the input signal and $E_1$, with $P_1$=16 mW and $P_2$=3 mW. The solid lines are the theoretically calculated spectral responses, as discussed in the text. Circles and squares are obtained with pulse durations of 6 µs and 3 µs, respectively.

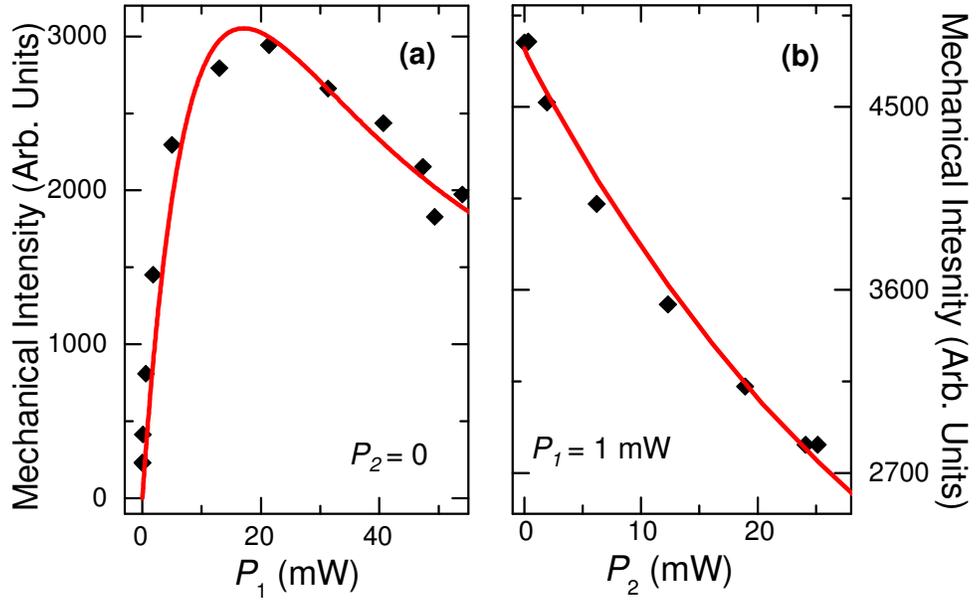

FIG 5. (color on line). (a) Intensity of the mechanical excitation as a function of $P_1$, with $P_2=0$. (b) Intensity of the mechanical excitation as a function of $P_2$, with $P_1=1$ mW. The solid lines are the result of a theoretical calculation, as discussed in the text.

Supplementary Materials for

# Optical wavelength conversion via optomechanical coupling in a silica resonator

Chunhua Dong[1], Victor Fiore[1], Mark C. Kuzyk[1], Lin Tian[2], and Hailin Wang[1]

[1]Department of Physics, University of Oregon, Eugene, Oregon 97403, USA
[2]Department of Physics, University of California, Merced, California 95343, USA

**1. Theoretical model**

We consider the optomechanical coupling between a mechanical oscillator and two optical cavity modes. The mechanical oscillator is coupled to the optical modes by two strong external laser fields near the respective red sideband (i.e., one mechanical frequency, $\omega_m$, below the cavity resonances, $\omega_1$ and $\omega_2$). In the resolved-sideband limit, the linearized optomechanical Hamiltonian is given by

$$H = \hbar\omega_m \hat{b}^+\hat{b} - \hbar(\Delta_1 \hat{a}_1^+ \hat{a}_1 + \Delta_2 \hat{a}_2^+ \hat{a}_2) + \hbar G_1(\hat{a}_1^+ b + \hat{a}_1 b^+) + \hbar G_2(\hat{a}_2^+ b + \hat{a}_2 b^+) \quad (1)$$

where $\hat{b}$ is the annihilation operator for the mechanical mode, $\hat{a}_1$ and $\hat{a}_2$ are the annihilation operators for the optical signal fields in the respective rotating frames of the external driving fields, $\Delta_1$ and $\Delta_2$ are the detuning between the external driving field and the relevant cavity resonance, and $G_1$ and $G_2$ are the effective optomechanical coupling rates between the respective optical signal field and the mechanical displacement.

With optical and mechanical damping processes included, the optomechanical coupling between the optical signal fields and the mechanical displacement can be described by the following equations of motion:

$$\dot{\alpha}_1 = [i(\omega_{in} - \omega_1) - \frac{\kappa_1}{2}]\alpha_1 - iG_1\beta + \sqrt{\kappa_1^{ext} I_{in}}$$

$$\dot{\alpha}_2 = [i(\omega_{in} - \omega_1 + \Delta_2 - \Delta_1) - \frac{\kappa_2}{2}]\alpha_2 - iG_2\beta \quad (2)$$

$$\dot{\beta} = [i(\omega_{in} - \omega_1 - \Delta_1 - \omega_m) - \frac{\gamma_m}{2}]\beta - i(G_1\alpha_1 + G_2\alpha_2)$$

where $\omega_{in}$ and $I_{in}$ are the frequency and photon flux of the input signal field for mode 1, respectively, $\kappa_i$ and $\kappa_i^{ext}$ ($i$=1, 2) are the total cavity decay rate and the effective output coupling rate of the two optical modes, respectively, $\gamma_m$ is the mechanical damping rate, and we define

$\alpha_1 = <\hat{a}_1> \exp[i(\omega_{in} - \omega_1 - \Delta_1)t]$, $\alpha_2 = <\hat{a}_2> \exp[i(\omega_{in} - \omega_1 - \Delta_1)t]$, $\beta = <\hat{b}> \exp[i(\omega_{in} - \omega_1 - \Delta_1)t]$.

The photon conversion efficiency in the steady state was obtained by solving Eq. 2 in the steady state. Suitable input and out relations were also used to obtain the output photon flux, $I_{out}$, for mode 2. The photon conversion efficiency is then given by

$$\chi = \frac{I_{out}}{I_{in}} = \eta_1 \eta_2 \frac{4 C_1 C_2}{(1 + C_1 + C_2)^2} \tag{3}$$

where $C_i = 4 G_i^2 / \gamma_m \kappa_i$ and $\eta_i = \kappa_i^{ext} / \kappa_i$ ($i$ = 1, 2) are the optomechanical cooperativity and the output coupling ratio, respectively. In addition, we also assumed that $\Delta_1 = \Delta_2 = -\omega_m$ and the input and output signal fields are resonant with the respective optical cavity modes.

The theoretical calculations shown in the figures of the main manuscript were obtained under the transient experimental conditions. The output photon flux was calculated according to the parameters used for the gated detection.

## 2. Experimental setup

Figure 1 shows a detailed schematic of the experimental setup for the optical wavelength conversion experiment. The input signal field and the driving field with λ ~ 800 nm were derived from a single-frequency Ti:Sapphire ring laser (Coherent 899-21). The driving field with λ ~ 637 nm was derived from a single-frequency dye ring laser (Coherent 899-21). The signal pulse and the two driving pulses were generated with three separate acousto-optic modulators (AOMs) and were then coupled into a single mode optical fiber. All three pulses were synchronized, with equal pulse duration. The input signal pulse double-passed its AOM such that we can tune the signal frequency by varying the AOM frequency without changing the propagating direction of the signal beam.

The two ring lasers were locked to the relevant whispering gallery modes (WGMs) of the silica resonator with a Pound-Drever-Hall (PDH) technique. For the PDH locking, we used the AOMs to generate 50 μs locking pulses that are temporally separate from the driving and signal pulses. The locking pulses were phase-modulated with an electro-optic modulator (EOM). A combination of lock-in amplifiers and PID controllers was used to generate the feedback signal. The integration time of the lock-in amplifiers was set to be 1 ms, which is slow compared with the repetition cycle of our experiments.

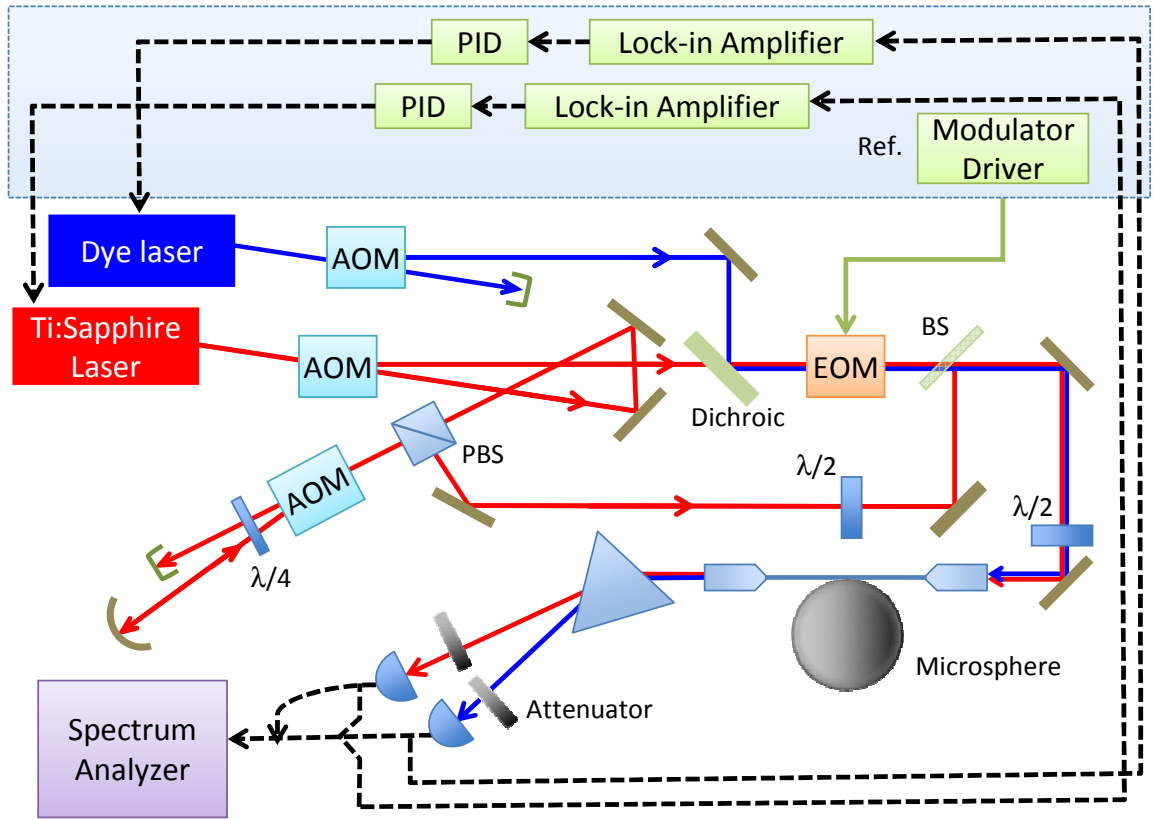

FIG. 1  The experimental setup, including the optical beam paths (solid lines) and electrical diagrams (dashed lines).

### 3. Gated-heterodyne-detection and calibration of the output signal

We used heterodyne detection to measure the output signal. For the heterodyne detection, the driving field with $\lambda \sim 637$ nm served as the local oscillator. For time-resolved measurements, we operated a spectrum analyzer in a time-gated detection mode and measured the power spectrum of the heterodyne signal as a function of the gate delay. Figure 2 shows a schematic of the gated detection. The data acquisition occurs only during the detection window. The time resolution of the measurement is limited by the resolution bandwidth as well as the gate length.

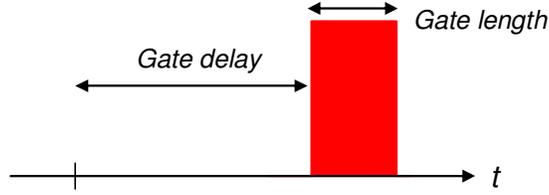

FIG. 2  Schematic for the gated detection.

Figure 3 shows an example of the power spectrum obtained with gated detection and under the experimental condition used for the data presented in the third figure of the main manuscript. The spectrally-integrated power density derived from the power spectrum measures the average power of the heterodyne signal within the detection window. We note that there is a thermal contribution to the output signal due to the Brownian motion of the mechanical oscillator. Figure 3 plots the thermal contribution obtained when the signal pulse was turned off. For a relatively strong coherent input signal field, the thermal contribution is small compared with that from the coherent mechanical excitation. With $C_1 \approx C_2 > 1$, the thermal contribution becomes negligible.

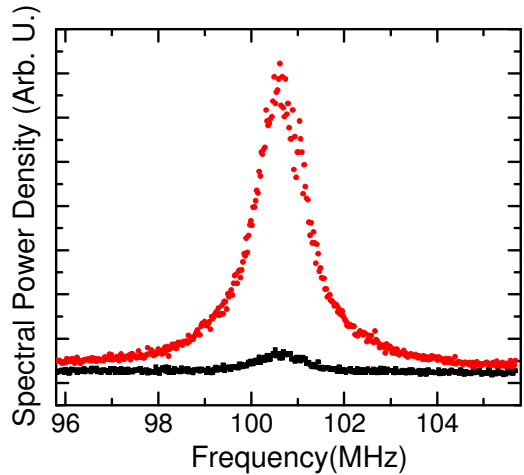

FIG. 3  A power spectrum of the heterodyne-detected output signal obtained with gated detection and under the experimental condition used for the data presented in the third figure of the main manuscript. The optical power of the driving fields with $\lambda \sim 800$ nm and $\lambda \sim 637$ nm are 1.7 mW and 2 mW, respectively. Red dots: input signal power = 0.2 mW. Blue dots: input signal power=0 mW.

We carried out separate calibration experiments, with the silica microsphere moved far away from the fiber taper, to relate the electrical power measured in the gated heterodyne detection to the optical power of the output signal. For the calibration, a laser pulse with the same frequency and duration as the output signal was generated with an AOM. The calibration pulse was then combined with a driving pulse, which serves as the local oscillator, and coupled into the single mode optical fiber. The optical powers of the calibration pulse and the local oscillator were measured separately from the output facet of the single mode fiber. (In order to obtain the wavelength conversion efficiency, the optical power of the input signal pulse was also measured at the output facet of the single mode fiber.) The power spectrum was then obtained with gated heterodyne detection under otherwise the same condition as the wavelength conversion experiment. For a given optical power of the local oscillator, this calibration relates directly the power of the heterodyne electrical signal to the optical power of the output signal. Note that care was taken such that the photo-detector always operated in a linear regime in both the calibration and the wavelength conversion experiments.

**4. Gated detection of the mechanical excitation**

We used a weak probe pulse at the red sideband of $\omega_1$ to probe the mechanical excitation involved in the state transfer between the optical and mechanical modes. As shown schematically in Fig. 4, the probe pulse is delayed relative to the driving pulse, $E_1$, which is also at the red sideband of $\omega_1$. The probe pulse is sufficiently weak such that it does not cause appreciable damping of the mechanical mode. Displacement power spectra were obtained with gated detection, with a gate length of 1 µs and with the gate at the center of the 3 µs probe pulse. The spectrally integrated area of the displacement power spectrum measures directly the intensity of the mechanical excitation.

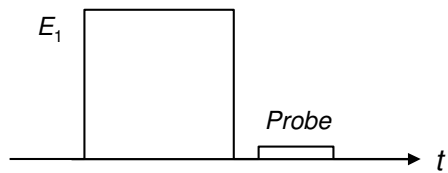

FIG. 4  The pulse sequence used for the gated detection of the mechanical excitation. The detection gate with a gate length of 1 µs is at the center of the 3 µs probe pulse.

To illustrate the effectiveness of the mechanical measurement, we plot in Fig. 5 the intensity of the mechanical excitation as a function of the gate delay. For this experiment, the coherent mechanical excitation was generated by $E_1$ and a synchronized signal pulse resonant with the optical mode. The mechanical lifetime obtained from the transient measurement is in general agreement with that derived from the displacement power spectrum obtained in a separate continuous-wave spectral domain measurement (without gated detection).

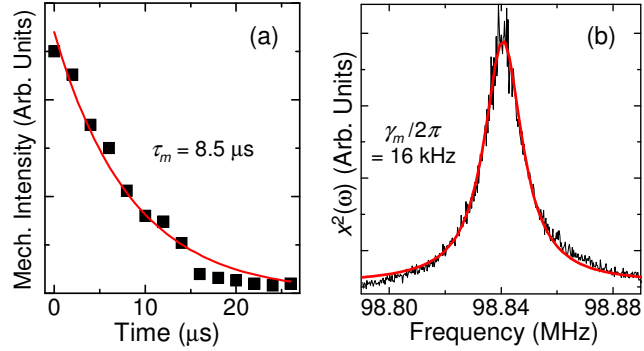

FIG. 5 (a) Mechanical intensity as a function of the gate delay. The solid red line is an exponential fit, yielding a mechanical lifetime of 8.5 µs. (b) Displacement power spectrum of the mechanical mode. The Lorenzian fit (red line) yields a linewidth of 16 kHz.